\documentclass{mem}
\usepackage{natbib}\usepackage{txfonts}\usepackage{balance}
\usepackage{graphicx}
\usepackage{hyperref}
\idline{75}{282}
\begin{document}
\def\teff{$T\rm_{eff }$}
\def\kms{$\mathrm {km s}^{-1}$}

\title{
Measurements of Nuclear
Reactions that Create and
Destroy Li and Be during BBN
}

   \subtitle{}

\author{B. \,Davids\inst{1,2}}

\institute{
TRIUMF, 4004 Wesbrook Mall, Vancouver, BC, V6T 2A3, Canada
\and
Department of Physics, Simon Fraser University, Burnaby, BC, Canada
\email{davids@triumf.ca}
}

\authorrunning{Davids}

\titlerunning{Reactions that Create and Destroy . . .}

\abstract{
I review measurements of the most important reactions involved in the creation and destruction of Li and Be during big bang nucleosynthesis (BBN) as well as their uncertainties and the relative contributions they make to the uncertainty in the primordial $^7$Li abundance ($^7$Li/H). Examining the sensitivity of calculated $^7$Li/H to these reactions as predicted by different BBN codes I find no significant differences. I compare my calculation of primordial $^7$Li/H to some recently published values and conclude that in the absence of a major undetected experimental blunder, nuclear physics uncertainties cannot account for the cosmological Li problem. With an estimated 13\% uncertainty in the calculated abundance, the discrepancy with observation amounts to some $4.6\sigma$. 
}
\maketitle{}

\section{Introduction}

All of the H, 90\% of the He, and around 1/4 of the Li observed in the universe are attributed to BBN \citep{johnson}. Standard BBN assumes the cosmological principle (that when viewed on sufficiently large scales the universe is everywhere isotropic and homogeneous), General Relativity, and the laws of physics discovered in terrestrial laboratories. Since the first modern BBN calculations were published more than 50 years ago, they have been continuously refined. The nuclear reaction network of \citet{wagoner} included only 144 reactions and decays, while that of \citet{coc} included 59 nuclides, 391 reactions, and 33 decays. Of these, neutron decay and 11 nuclear reactions have the largest influence on the primordial $^2$H, $^{3,4}$He, and $^7$Li abundances \citep{smith}.

Wagoner's original code NUC123 was ported to VAX FORTRAN by \citet{kawano}. Other public codes include PArthENoPE \citep{pisanti}, AlterBBN \citep{arbey}, and PRIMAT \citep{pitrou}. I have modified the Wagoner/Kawano code by updating the rates of nuclear reactions relevant to the production and destruction of Li in the early universe. The abundances calculated by BBN codes depend strongly on three parameters: the mean lifetime of the neutron $\tau_n$, the number of light neutrino flavours $N_{\nu}$, and the current baryon-to-photon ratio $\eta_0$. In the calculations performed with my version of the Wagoner/Kawano code and PRIMAT described here, the values of  $\tau_n = 879.5(8)$~s \citep{serebrov}, $N_{\nu}=3$, and $\eta_0=6.091(44)\times10^{-10}$ \citep{ade} were adopted. As my focus is on the cosmological Li problem \citep{fields}, I estimated the uncertainty in its calculated primordial abundance but have not quantified those of the other light elements.

\section{Results}
If $\eta_0$ were small, $^7$Li would principally be produced directly via $t(\alpha,\gamma)^7$Li, but at the value of $\eta_0$ inferred from cosmic microwave background observations, it is mainly produced indirectly via $^3$He$(\alpha,\gamma)^7$Be with subsequent electron capture to $^7$Li following recombination. The destruction of $^7$Li during BBN proceeds dominantly via $^7$Li($p,\alpha)\alpha$, whereas $^7$Be is destroyed principally by way of  $^7$Be$(n,p)^7$Li($p,\alpha)\alpha$.

\begin{figure}[]
\resizebox{\hsize}{!}{\includegraphics[clip=true]{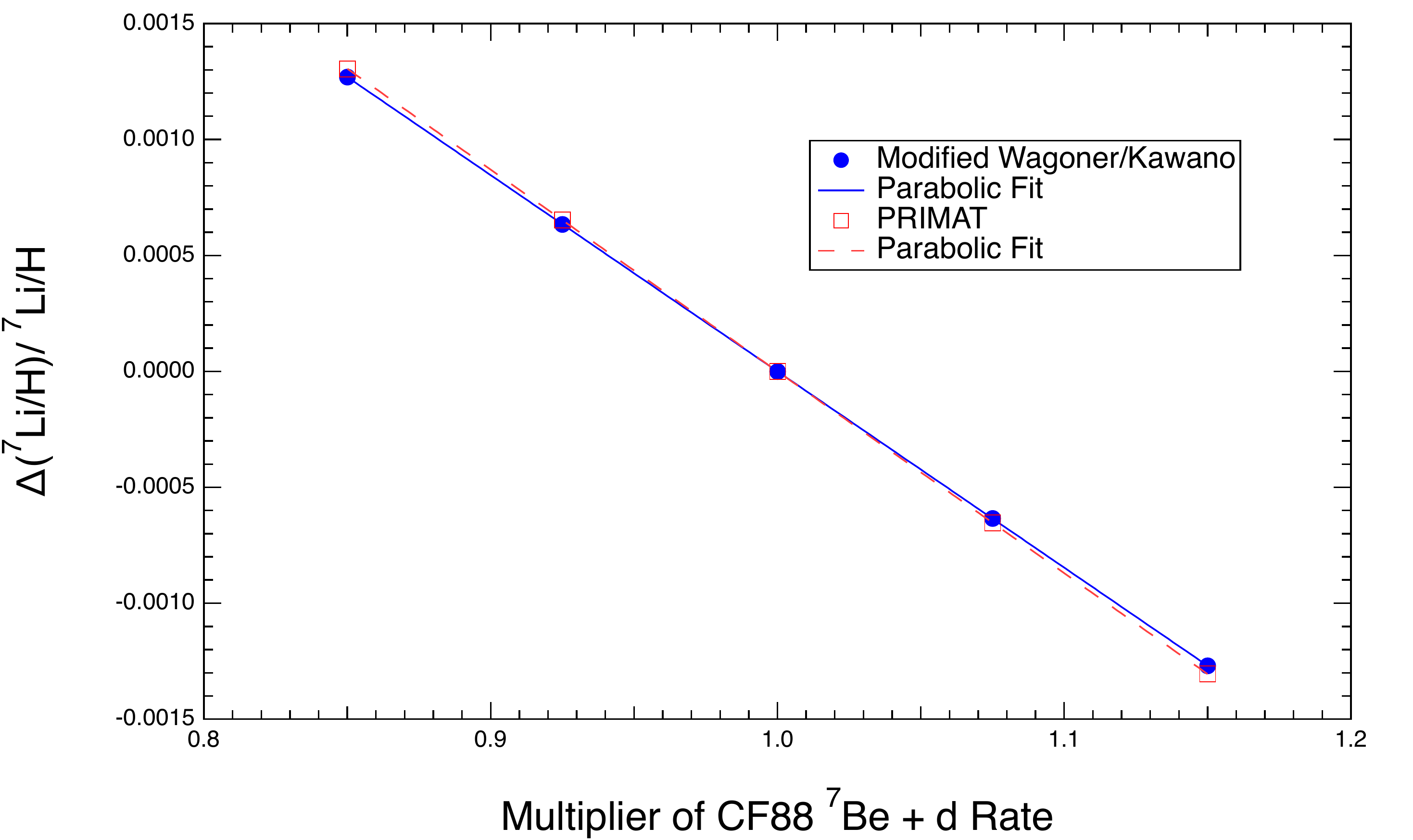}}
\caption{
\footnotesize
Relative variation of primordial $^7$Li/H as a function of the factor multiplying the \citet{caughlan} estimate of the $d+^7$Be reaction rate. Calculations performed with both PRIMAT and a modified version of the Wagoner/Kawano code were fit with quadratic polynomials and the consistent sensitivities shown in Table \ref{sens} obtained.
}
\label{sensfig}
\end{figure}

To quantify the dependence of calculated primordial abundances on nuclear reaction rates, we employ the sensitivity $\alpha_n$ defined by \citet{fiorentini} that can be obtained from the logarithmic derivative of an abundance or mass fraction $X$ with respect to a parameter $p_n$ through \begin{equation}
X=X_{0}\prod_n(\frac{p_n}{p_{n,0}})^{\alpha_n}.
\end{equation} The sensitivities of calculated primordial $^7$Li/H to important reaction rates have been tabulated by \citet{cyburt,cocvangioni,cfoy}. BBN calculations reported by the authors of a recent $d(^7$Be,$\alpha)$ measurement \citep{rijal}, if correct, would imply that the sensitivity of $^7$Li/H to this rate is large. I calculated it using my modified Wagoner/Kawano code and PRIMAT. Adopting the \citet{caughlan} (CF88) estimate of the $d+^7$Be rate, I varied it by $\pm7.5\%$ and $\pm15\%$, studying the relative change in primordial $^7$Li/H as a function of the factor by which the CF88 rate is multiplied, and fitting with quadratic polynomials. The fits are shown in Fig.\ \ref{sensfig} and the results appear in Table \ref{sens}, which demonstrates that the sensitivities calculated with these codes agree and that the calculated primordial $^7$Li/H is almost entirely insensitive to the rate of $d(^7$Be,$\alpha)$.

\begin{table*}
\begin{center}
\caption{Calculated Sensitivities of Primordial $^7$Li/H to Nuclear Reaction Rates}
\label{sens}
\begin{tabular}{ccccc}
\hline
Reaction & \citet{cocvangioni} & \citet{cfoy} & Modified Wagoner/Kawano &PRIMAT\\
\hline
$n(p,\gamma)d$ & 1.33 & 1.339 & & \\
$d(p,\gamma)^3$He & 0.57 & 0.589 & & \\
$d(d,n)^3$He & 0.69 & 0.698 & & \\
$^3$He($d,p)\alpha$ & -0.75 & -0.752 & & \\
$^3$He($\alpha,\gamma)^7$Be & 0.97 & 0.963 & & \\
$^7$Be($n,p)^7$Li & -0.71 & -0.705 & & \\
$d(^7$Be,$\alpha)p \alpha$ & & & -0.0085 & -0.0087 \\
\hline
\end{tabular}
\end{center}
\end{table*}

Despite agreement on the sensitivities, the primordial Li abundances calculated by the codes differ on account of the adopted rates of the important nuclear reactions. The sources of the rates of the most important nuclear reactions for Li production adopted by the three codes compared here are summarized in Table \ref{sources}. Below I discuss the uncertainties in these reactions and their contributions to the calculated primordial Li abundance.

\begin{table*}
\begin{center}
\caption{Literature Sources of Nuclear Reaction Rates Important for Primordial $^7$Li}
\label{sources}
\begin{tabular}{cccc}
\hline
Reaction & Modified Wagoner/Kawano & \citet{cfoy} & \citet{pitrou}\\
\hline
$n(p,\gamma)d$ & \citet{ando} & \citet{ando} & \citet{ando} \\
$d(p,\gamma)^3$He & \citet{cyburt} & \citet{nacre} &  \citet{iliadis} \\
$d(d,n)^3$He & \citet{cyburt} & \citet{nacre} & \citet{gomez} \\
$^3$He($d,p)\alpha$ & \citet{cyburt} & \citet{nacre} & \citet{descouvemont} \\
$^3$He($\alpha,\gamma)^7$Be & \citet{cyburtdavids} & \citet{cyburtdavids} & \citet{iliadis} \\
$^7$Be($n,p)^7$Li & \citet{cyburt} & \citet{cyburt} & \citet{descouvemont} \\
\hline
\end{tabular}
\end{center}
\end{table*}

The primordial $^7$Li abundance is, somewhat surprisingly among reactions, most sensitive to the $n(p,\gamma)d$ rate due to its effect on the number of neutrons available to destroy the $^7$Li progenitor $^7$Be after its formation. The effective field theory calculation of \citet{ando} of the $n(p,\gamma)d$ reaction has been generally adopted. Its Markov chain Monte Carlo analysis includes the low energy capture data of \citet{suzuki,nagai,tornow,schreiber} and the photodisintegration data of \citet{moreh} and \citet{hara}. The estimated uncertainty is 1\%, implying an associated uncertainty in $^7$Li/H of 1.3\%.

While critical for D abundance predictions, the $d(p,\gamma)^3$He rate strongly influences $^7$Li as well. Its $S$ factor was evaluated by \citet{iliadis}, who relied upon the low energy data of \citet{ma,schmid,casella,bystritsky} and the {\it ab initio} calculation of \citet{marcucci}. This approach resulted in an estimated 3.7\% uncertainty in the cross section at the energies below 200 keV most relevant to BBN. The \citet{adelberger} evaluation of the same data excepting \citet{bystritsky} employed a quadratic polynomial fit with a different shape from the calculation of \citet{marcucci} and obtained a slightly lower $S(0)$ with an estimated uncertainty of (+7.9\%,-7.5\%). The polynomial fit of \citet{cyburt} to $S(E)$ used here is consistent with both of these and has an estimated 6.3\% uncertainty, implying an associated uncertainty in $^7$Li/H of 3.7\%.

The $d(d,n)^3$He rate must be known at approximately the same temperatures as $d(p,\gamma)^3$He. \citet{cocpetitjean} used the {\it ab initio} model of \citet{arai} to fit the data of \citet{leo,gre,bro,kra} below 600 keV supplemented with an R-matrix calculation at higher energies. The estimated uncertainty is 1.1\%, which is to be compared with the 5.4\% estimate of \citet{cyburt}, leading to 0.8\% and 3.8\% uncertainties in primordial $^7$Li/H, respectively.

The cross section of $^3$He($\alpha,\gamma)^7$Be must be known well at energies between 260 and 580 keV. Modern prompt- and delayed-$\gamma$-ray data and recoil separator measurements of \citet{singh,bemmerer,confortola,brown,gyurky,dileva} have been understood with the aid of microscopic models of \citet{nollett} and \citet{neff}. Consistent evaluations of these low energy data by \citet{cyburtdavids,adelberger,iliadis} resulted in uncertainty estimates of 7.4\%, 5.1\%, and 2.1\%, respectively, implying associated primordial $^7$Li/H uncertainties of 2.0-7.2\%.

Since it is responsible for destroying $^3$He before it can capture an $\alpha$ particle to form $^7$Be, the primordial Li abundance is anti-correlated with the $^3$He($d,p)\alpha$ cross section in the BBN energy range from circa 60-360 keV. The polynomial fit of \citet{cyburt} to the data of \citet{bonner,kunz,zhichang,moller,kra} and \citet{geist} resulted in a 6.6\% rate uncertainty, translating into a 5.0\% uncertainty in primordial $^7$Li/H.

As the primary destruction channel, $^7$Be($n,p)^7$Li is a reaction whose rate must be well known, but as both reactants are radioactive its measurement presents special challenges. A recent CERN measurement \citep{damone} at neutron energies $\leq325$ keV found a cross section significantly higher than that of \citet{koehler}, who only covered energies below 13.5 keV. \citet{damone} estimate an uncertainty of circa 11\% in their rate at BBN temperatures, which also relies upon the $^7$Li($p,n)^7$Be data of \citet{sekharan}. Compared to BBN calculations performed with the \citet{cyburt} rate, with its estimated 4.8\% uncertainty, the \citet{damone} rate results in a 4.4\% decrease in primordial Li. In my Wagoner/Kawano code calculations I assume an 11\% uncertainty, implying an associated 8\% uncertainty in primordial $^7$Li.

While $^7$Be($n,\alpha)\alpha$ was not thought to be a particularly significant destruction reaction, a reanalysis of $\alpha(\alpha,n)^7$Be and $\alpha(\alpha,p)^7$Li data by \citet{hou} and a direct measurement by \citet{barbagallo} are consistent in finding that \citet{wagoner} overestimated the rate. Using the experimentally determined rate increases $^7$Li/H by 1\%.

Despite the insensitivity of $^7$Li/H to the $d+^7$Be rate, for completeness I now consider recent work on this reaction. After the Wilkinson Microwave Anisotropy Probe determined $\eta_0$ with high precision, \citet{cocangulo} recalculated BBN light element abundances and proposed that the discrepancy between the calculated and observationally inferred $^7$Li abundance could be resolved if the low energy cross section of $d(^7$Be,$p)2 \alpha$ were $\geq100$ times as large as the commonly used CF88 rate, which was based on the estimate of \citep{parker}. This conjecture inspired a new measurement of the $d(^7$Be,$p)2 \alpha$ cross section that found its cross section to be smaller than Parker's estimate and it could not resolve the cosmological Li problem \citep{angulo}.

\citet{pospelov} examined the possibility of resonant enhancement of several $^7$Be destruction reactions during BBN, including $d(^7$Be,$p)2 \alpha$, and defined the range of energies and partial widths of a hypothetical state in $^9$B that could provide the requisite enhancement. They found that a resonance at $d+^7$Be energies between 170 and 220 keV having a partial deuteron width $\Gamma_d$ between 10 and 40 keV and a comparable exit width $\Gamma-\Gamma_d$ could resolve the cosmological Li problem. If such a state decayed via proton emission through the 16.626 MeV state in $^8$Be, it would not have been observed in the \citet{angulo} measurement, as it was not.

Seeking a $^9$B state with these properties, Kirsebom found that a candidate state's energy and width had been measured by \citet{scholl}. \citet{kirsebom} showed that the 16.800(10) MeV state in $^9$B reported by \citet{scholl} could not resolve the Li problem, even under the most optimistic assumption about its partial widths, finding a maximum reduction in $^7$Li/H of 3.5(8)\%. The authors noted that their result held for both $d(^7$Be,$p)2 \alpha$ and $d(^7$Be,$\alpha)p \alpha$ and that this state could have very little effect on the primordial $^7$Li abundance through resonant enhancement of $d+^7$Be, regardless of the exit channel.

\begin{figure}[]
\resizebox{\hsize}{!}{\includegraphics[clip=true]{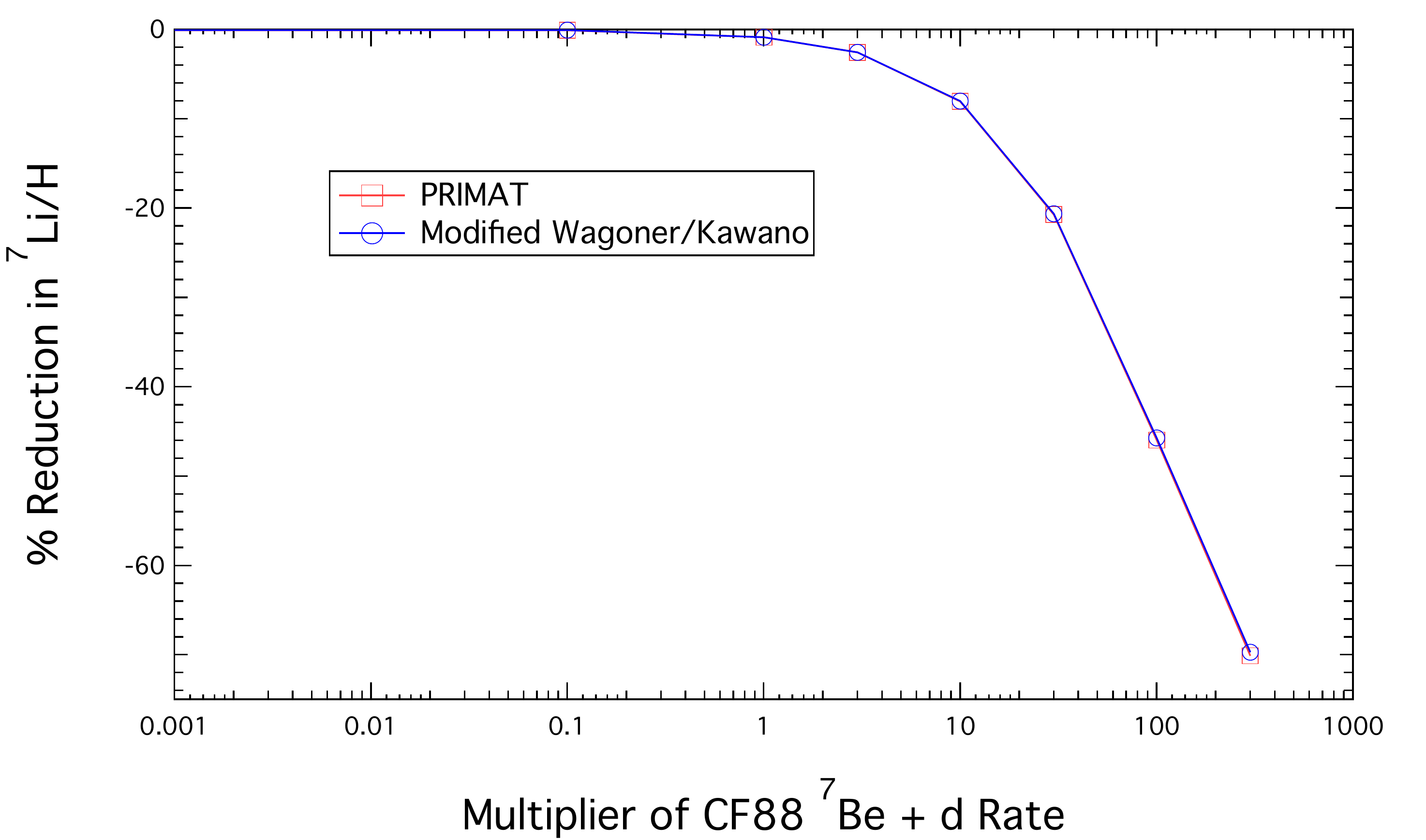}}
\caption{
\footnotesize
Reduction of primordial $^7$Li/H as a function of the factor multiplying the \citet{caughlan} estimate of the $d+^7$Be reaction rate compared to the result neglecting this reaction entirely. Calculations performed with PRIMAT and my modified Wagoner/Kawano code are consistent.
}
\label{litheory}
\end{figure}

Based on a direct measurement of $d+^7$Be, \citet{rijal} reported a resonance at 360(50) keV with a strength of 1.7(5) keV consistent with the \citet{scholl} state at 310(11) keV. Based on the high energy and weak resonance strength, such a state would result in even less resonant enhancement of the $d+^7$Be rate than that found by \citet{kirsebom}. Indeed, as noted by \citet{gai} and \citet{cocdavids}, the \citet{rijal} rate is practically identical to that of CF88 above 0.6 GK, dropping smoothly to about 1/2 of the CF88 rate at 0.1 GK. Unfortunately, the BBN calculations of \citet{rijal} rely upon an inadequately tested BBN code whose results are manifestly inconsistent with those of the Wagoner/Kawano and PRIMAT codes. As shown in Fig.\ \ref{sens} and Fig.\ \ref{litheory}, these well-tested codes are quantitatively consistent regarding the sensitivity of calculated primordial $^7$Li/H to both large and small changes in the $d+^7$Be rate. \citet{fieldsolive} report similar findings.

\begin{figure}[]
\resizebox{\hsize}{!}{\includegraphics[clip=true]{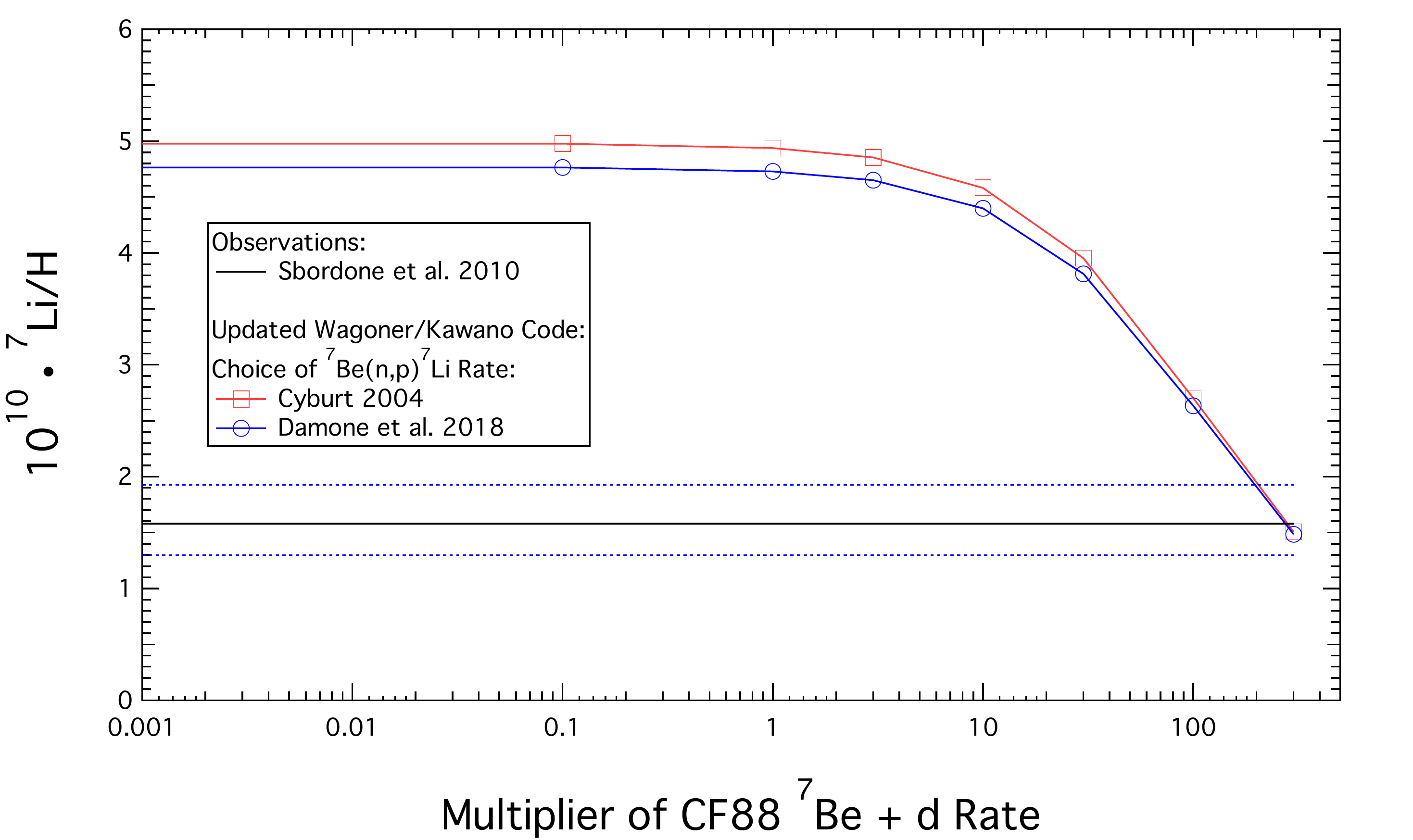}}
\caption{
\footnotesize
Primordial $^7$Li/H as a function of the factor multiplying the \citet{caughlan} estimate of the $d+^7$Be reaction rate. Modified Wagoner/Kawano code calculations performed alternatively assuming the \citet{cyburt} and \cite{damone} $^7$Be($n,p)^7$Li rates are shown along with observations of \citet{sbordone}.
}
\label{obs}
\end{figure}

Abundance calculations carried out with my modified Wagoner/Kawano code are compared with those of \citet{cfoy} and \citet{pitrou} in Table \ref{abun}. The light element abundances calculated by the different codes are in good agreement with each other but $^7$Li is grossly discrepant with observations. Fig.\ \ref{obs} plots primordial $^7$Li/H as a function of a factor multiplying the CF88 estimate of the $d+^7$Be reaction rate calculated alternatively with the \citet{cyburt} or \citet{damone} $^7$Be($n,p)^7$Li rate. The disagreement between standard calculations of $^7$Li/H with that inferred from observations \citep{sbordone} is stark. I estimate a 13\% uncertainty in the calculated primordial $^7$Li/H, which is discrepant with observations by some $4.6\sigma$.

\begin{table*}
\begin{center}
\caption{Calculated Primordial Light Element Abundances}
\label{abun}
\begin{tabular}{cccc}
\hline
Quantity & Modified Wagoner/Kawano & \citet{cfoy} & \citet{pitrou}\\
\hline
$\textrm{Y}_P$  & 0.247 & 0.24709(25) & 0.24709(17)\\
$\textrm{D/H}\times10^5$ & 2.57 & 2.58(13) & 2.459(36)\\
$^3\textrm{He/H}\times10^5$ & 1.00 & 1.0039(90) & 1.074(26)\\
$^7\textrm{Li/H}\times10^{10}$ & 4.94(65) & 4.68(67) & 5.623(247)\\
\hline
\end{tabular}
\end{center}
\end{table*}

\section{Conclusions}
 
Barring the existence of one or more undetected major experimental blunders, recent cross section measurements have confirmed that nuclear reaction rate uncertainties cannot account for the cosmological Li problem. Well tested BBN codes agree on the sensitivities to the reaction rates and yield consistent primordial abundances.
 
\begin{acknowledgements}
I gratefully acknowledge helpful discussions with Alain Coc and Cyril Pitrou and generous support from the NSERC of Canada.
\end{acknowledgements}

\bibliographystyle{aa}


\end{document}